\documentclass[aps,pra,twocolumn,twoside,floatfix,superscriptaddress,showpacs]{revtex4}
\usepackage{hyperref}
\usepackage{bbm}
\usepackage{amsmath, amssymb}
\usepackage{color}
\usepackage{graphicx,epsfig}
\usepackage{pifont}
\usepackage{subfigure}
\usepackage{amssymb}
\usepackage{amsmath}
\usepackage{amssymb}
\usepackage{graphicx}
\usepackage{subfigure}
\usepackage{lscape}
\graphicspath{{./picture/}}

\begin{document} 
\title{Entanglement of distant optomechanical systems}
\author{C. Joshi}
\email{cj53@hw.ac.uk}
\affiliation{SUPA, Department of Physics, Heriot-Watt University, Edinburgh, EH14 4AS, UK}
\author{J. Larson}
\affiliation{Department of Physics, Stockholm University, AlbaNova University Center, 106 91 Stockholm, Sweden}
\affiliation{Institut f\"ur Theoretische Physik, Universit\"at zu K\"oln, K\"oln, 50937, Germany}
\author{M. Jonson}
\affiliation{SUPA, Department of Physics, Heriot-Watt University, Edinburgh, EH14 4AS, UK}
\affiliation{Department of Physics, University of Gothenburg, SE-412 96 G{\"o}teborg, Sweden}
\affiliation{Department of Physics, Division of Quantum Phases \& Devices, Konkuk University, Seoul 143-701, Korea}
\author{E. Andersson}
\affiliation{SUPA,  Department of Physics, Heriot-Watt University, Edinburgh, EH14 4AS, UK}
\author{P. \"Ohberg}
\affiliation{SUPA,  Department of Physics, Heriot-Watt University, Edinburgh, EH14 4AS, UK}

\begin{abstract} 
We theoretically investigate the possibility 
to generate non-classical states of optical and mechanical modes of optical cavities, distant from each other. A setup comprised of two identical cavities, each 
 with one fixed and one movable mirror and coupled by an optical fiber, is studied in detail. We show that with such a setup there is potential to generate entanglement between the distant cavities, involving both optical and mechanical modes. The scheme is robust with respect to dissipation, and nonlocal correlations are found to exist in the steady state at finite temperatures. 
\end{abstract} 
\maketitle
\noindent

\section{Introduction}
\label{sec:introdu}
Quantum entanglement is 
one of the most intriguing features of
quantum mechanics. Although quantum mechanics has proven to be highly 
successful in explaining 
physics at 
microscopic and subatomic scales, its validity at macroscopic or even mesoscopic scales is still debated. Some of the astonishing features appear when we try to apply 
quantum mechanical principles to  macroscopic systems. Superpositions of macroscopic systems is one  example~\cite{schr}.

It is not yet completely clear to what extent quantum mechanics applies to macroscopic objects. Quantum phenomena such as entanglement generally do not appear in the macroscopic world. The difficulty of seeing quantum superpositions of macroscopic systems is often attributed to environment-induced decoherence. Such decoherence is thought to be the main cause reducing any quantum superposition to a classical statistical mixture \cite{zure}.  Thus, an obvious but impractical choice would be to minimize the detrimental effect of the environment through perfect isolation of  the system of interest. Nonetheless, with the spectacular level of experimental advancements,  the possibility of seeing macroscopic quantum superpositions  appears to be within reach~\cite{c60}. 

Related to this, quantum engineering~\cite{qcon} in the field of optomechanics has made rapid advancement~\cite{flor}.  In a typical optomechanical setup, a mechanical system can be manipulated by radiation forces. Such systems have recently attracted much theoretical and experimental attention~\cite{tjki,tjki2}. This is partly because of their potential usefulness in extremely sensitive sensor technology and in quantum information processing~\cite{tjki}. Also, they are potentially one of the best tools to test fundamentals of quantum mechanics. 
Seminal progress has been made both theoretically and experimentally in 
this novel emerging field~\cite{flor,tjki2}. 

 In a 
typical setting using  optomechanical interaction, the main component is a cavity with a movable mirror. Light in the cavity and the movable mirror interact due to a coupling induced by the radiation pressure. As a result, the movable mirror executes harmonic motion around its equilibrium value~\cite{lasing}, which thus alters the cavity resonance frequency. This, in turn, changes the circulating power in the cavity and hence the radiation pressure force acting on the movable mirror, leading to intrinsic nonlinearities~\cite{walther}. Strong light-matter coupling, both for opto- and for electro-mechanical systems, is a main ingredient in this emerging research field~\cite{simo,teuf}. Within the strong coupling regime, radiation-pressure interaction has been successfully utilized for
ground state cooling of mechanical oscillators~\cite{iwra,simo3,connell}. Some of the fascinating schemes include preparing the cavity mode and the  movable mirror in a non-classical state \cite{smans,sbos},  preparing  optomechanical or fully mechanical Schr\"odinger cat states ~\cite{smans3,agarwal,laura}, and even inducing quantum correlations between the 
subsystems~\cite{shbar,mment,mfent,mment2}.  Apart from the  mostly studied cavity-movable mirror geometry, there have been some recent breakthroughs in  exploring quantum features  of a membrane in a cavity~\cite{jdthom, bhatta}. There are also recent proposals exploring the possibility of observing  photonic analogs of the Josephson effect in an optomechanical setting~\cite{larson}.

A common feature of most of these studies involve  enhancement of the radiation pressure coupling through intense laser driving of the cavity field. This is required to achieve strong radiation pressure coupling which otherwise is too weak to observe any non-classical phenomenon. Although  most of these studies are restricted to Gaussian state preparation involving optomechanical interaction, there have been some recent proposals to study non-Gaussian quantum states in the regime of  single-photon optomechanics \cite{akram,anunn}.

Motivated by these theoretical and experimental advancements,  we shall here explore 
the possibility of entangling mechanical and optical modes of two distant cavities. In previous studies of the entanglement of 
distant mechanical mirrors, squeezed light 
was used as an available resource in order to entangle the two distant mirrors, which 
were either part of the same optical cavity~\cite{agarwal,mment} or belonged to two different cavities~\cite{laura}.  In the present work we are interested in a physically different setup, where  two distant Fabry-P\'erot cavities each fabricated with one movable mirror are coupled by
an optical fiber.
We show that, as a result of a combination of the optomechanical interaction and an optical-fiber mediated coupling, the two distant optical and mechanical modes become entangled.  Moreover, 
we explicitly study two different regimes of physical interest. First we 
will impose an approximation of Born-Oppenheimer (BO) type, and consider a scenario in which the two cavities are not externally pumped. In this regime the two mechanical modes are found not to be very strongly entangled. The advantage is however that the approximation allows us to find an analytical solution describing the evolution of the state of the two mirrors. Thereafter we work in a regime where the coupled cavities are strongly driven. Here we explicitly derive the relevant quantum Langevin equations (QLE) and  construct the covariance matrix governing the dynamics of all the optical and mechanical modes. 
 
The paper is organized in the following way. 
Section \ref{sec:thery} introduces 
the theoretical model and the physical setup.  This forms the framework for Sec.
\ref{sec:unitry} in which the unitary evolution of the system is studied, followed by 
analysis of the dissipative regime in Sec. \ref{sec:disip} and a brief discussion on the validity of the BO approximation in Sec. \ref{sec:BOvalidity}. A quantum Langevin approach is introduced in
Sec. \ref{sec:langv}, and  
we conclude with a short discussion in Sec. \ref{sec:conclude}. 
 
\section{Physical setup}
\label{sec:thery}
\begin{figure}[!]
\centering
\includegraphics[width=0.451\textwidth]{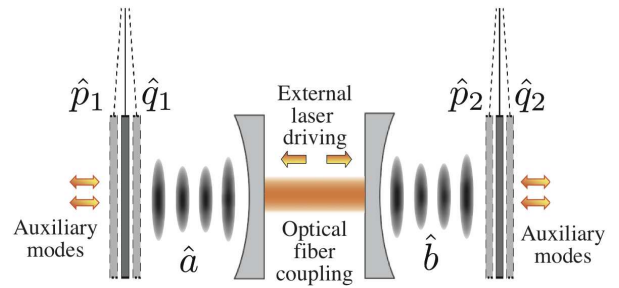}
\caption{\label{figzero}(Color online) Sketch of the physical setup to entangle  distant optomechanical modes.  Two
optomechanical cavities pumped by classical laser fields are coupled to each other  by an optical fibre. As a result of  indirect coupling mediated by the two cavity modes, the two movable mirrors become entangled. Furthermore, two initially uncorrelated auxiliary cavity modes interact independently with the two entangled movable mirrors, which induces non-local correlations between the two modes. Using standard homodyne measurement techniques non-local correlations between the two auxiliary cavity modes can be read out giving an indirect signature of quantum correlations between the two mirrors.}
\end{figure}

We consider a physical setup comprised  of two identical Fabry-P\'erot cavities, each 
with one fixed and one movable mirror, as schematically shown in  Fig.~\ref{figzero}. We assume that only one resonant mode of each cavity is populated, and that these two modes are coupled via an optical fiber.  
The two modes have the same frequency, $\omega= 2\pi c  /L$; where $L$ is the cavity length, and are described by the creation (annihilation) operators $ \hat{a}^{\dagger}(\hat{a})$ and $ \hat{b}^{\dagger}(\hat{b})$, respectively.
Furthermore, we assume that each movable mirror has been cooled near to its ground state, so that it is operating in the quantum regime. Under the action of cavity-photon-induced radiation pressure, the movable mirrors will oscillate about their equilibrium positions. 

If we assume that the  two mirrors move distances $x$ and $y$ along the respective cavity axes, so that the two displacements are much smaller than the wavelength of each cavity mode in one cavity round-trip time, then scattering of photons to other cavity modes can be safely neglected \cite{afpa,law}. 
The effective lengths of the cavities will then  become $L+x$ and $L+y$, with  new resonance frequencies   $\omega_a=  2\pi c  /(L+x)$ and  $\omega_b=  2\pi c  /(L+y)$, where $x$ and $y$ are the instantaneous displacements of the two cavity mirrors from their equilibrium positions. 
With the above assumption, i.e. 
$x/L, y/L\ll1$, 
the free evolution of the two optical cavity modes 
in the adiabatic regime takes the form~\cite{law}
 \begin{eqnarray}
\hat{H}_{\rm free}&=&\hbar \omega_{a} \hat{a}^{\dagger}\hat{a}+\hbar \omega_{b}\hat{b}^{\dagger}\hat{b}\\ \nonumber
&=&\hbar \omega\left(1+\frac{x}{L}\right)^{-1} \hat{a}^{\dagger}\hat{a}+\hbar \omega\left(1+\frac{y}{L}\right)^{-1} \hat{b}^{\dagger}\hat{b}\\ \nonumber
&\approx& \hbar \omega \left(\hat{a}^{\dagger}\hat{a}+ \hat{b}^{\dagger}\hat{b}\right)- \frac{\hbar \omega}{L} \hat{a}^{\dagger}\hat{a} x- \frac{\hbar \omega}{L} \hat{b}^{\dagger}\hat{b} y.
\end{eqnarray}
Under the action of a weak radiation-pressure force,  each movable mirror undergoes small-amplitude oscillations with frequency $\Omega$. In the absence of 
external driving,  the full Hamiltonian of the two coupled cavities thus  becomes
\begin{equation}\label{nayihm}
\hat{H}=\hat{H}_{\rm free}+ \frac{m\Omega ^{2}}{2} x^{2}+\frac{p_{x}^{2}}{2m}+ \frac{m\Omega ^{2}}{2} y^{2}+\frac{p_{y}^{2}}{2m}+\hbar\lambda\left(\hat{a}^{\dagger}\hat{b}+\hat{b}^{\dagger}\hat{a}\right),
\end{equation}
where $\lambda$ is the inter-mode coupling between the two cavities. This coupling could be mediated by, e.g., an optical fiber connecting the two distant cavities. Introducing dimensionless conjugate variables $q_{i}$ and $p_{i}$  for the $i$th movable mirror, \eqref{nayihm} can be rewritten as 
\begin{eqnarray}\label{firstham}
\hat{H} & = & {\hbar} 
{\omega\left(\hat{a}^{\dagger}\hat{a}+\hat{b}^{\dagger}\hat{b}\right)+\frac{\hbar\Omega}{2}\left(\hat{q_{1}}^{2}+\hat{p_{1}}^{2}\right)+\frac{\hbar\Omega}{2}\left(\hat{q_{2}}^{2}+\hat{p_{2}}^{2}\right)}\nonumber \\   
& & +\hbar\lambda\left(\hat{a}^{\dagger}\hat{b}+\hat{b}^{\dagger}\hat{a}\right)
-\hbar g\left(\hat{a}^{\dagger}\hat{a}\hat{q_{1}}+\hat{b}^{\dagger}\hat{b}\hat{q_{2}}\right), 
\end{eqnarray}
where $g= (\omega/ L) \sqrt{\hbar/ m \Omega}$ is the radiation pressure-induced coupling between the cavity modes and the movable mirrors. 
The Hamiltonian (\ref{firstham}) will form the basis for the analysis in the next section, where we will make an adiabatic approximation that allows us to 
study the unitary evolution of the two movable mirrors. 

\section{Born-Oppenheimer  approach}
\label{sec:born}
\subsection{Effective adiabatic model}\label{ssec:admod}
The frequency mismatch between optical ($\omega/2\pi \sim 10^{14}$~Hz) and mechanical ($\Omega/2\pi \sim 10^{6} -10^{9}$~Hz)  degrees of freedom is enormous~\cite{tjki2}. This suggests a separation of the Hamiltonian \eqref{firstham} into two parts,  one with very 
rapidly  evolving  optical modes and another with slowly varying mechanical modes.
In the limit that the mirror coordinates $\hat{q}_{1}$ and  $\hat{q}_{2}$ remain stationary with respect to the 
rapidly evolving cavity modes $\hat{a}$ and $\hat{b}$, we can diagonalize the interaction between the two cavity modes of Hamiltonian  \eqref{firstham}. 

We first introduce the collective excitation operators $\hat{A}$ and $\hat{B}$ obeying
\begin{equation}\label{rotan}
\left(
\begin{array}{c}
\hat{a} \\
   \hat{b}  
\end{array}
\right)=\left(
\begin{array}{cc}
\rm{cos}\theta& \rm{sin}\theta \\
  -\rm{sin}\theta &\rm{cos}\theta 
\end{array}
\right)\left(
\begin{array}{c}
\hat{A} \\
   \hat{B}  
\end{array}
\right).
\end{equation}
Choosing $\tan 2 \theta=2 \lambda/ (g(q_{1}-q_{2}))$ and substituting for the new field modes, the rapidly  varying optical  part of Hamiltonian \eqref{firstham} reduces to 
\begin{eqnarray}\label{digham}
\hat{H}_{\rm{cav}} & = & \hbar \left (\omega-g\frac{q_{1}+q_{2}}{2}\right) (\hat{A}^{\dagger}\hat{A}+\hat{B}^{\dagger}\hat{B}) \nonumber \\
 & & - \hbar \sqrt{g^{2}{\left (q_{1}-q_{2}\right)^{2}+4\lambda^{2}}} \frac{(\hat{A}^{\dagger}\hat{A}-\hat{B}^{\dagger}\hat{B})}{2}. 
\end{eqnarray}
The Hamiltonian describing the dynamics of the two movable cavity mirrors thus takes the form
\begin{eqnarray}\label{mirham}
\hat{ H}_{\rm{mir}}=\hbar \frac{\Omega}{2}\left(\hat{q_{1}}^{2}+\hat{p_{1}}^{2}+\hat{q_{2}}^{2}+\hat{p_{2}}^{2}\right)+\hat{H}_{\rm{cav}}. 
\end{eqnarray}
The treatment this far is exact. Typically, the cavity field will adiabatically follow the slow motion of the two movable mirrors. Thus  by considering rapidly varying collective cavity  modes and slowly varying mirror modes we can make
the  BO approximation~\cite{BO,lsced}, and write the collective wave function of the cavity-mirror coupled system at time $t$ as
\begin{equation}\label{mirham1}
|\Psi(t) \rangle=\sum_{n_{A},n_{B}}P(n_{A},n_{B}) |n \rangle |\Phi(n,t) \rangle. 
\end{equation}
Here, $P(n_{A},n_{B})$ is the 
probability distribution of the collective cavity fields,  $|n \rangle$=$|n_{A},n_{B} \rangle$ denotes the time-independent index of the energy levels of the two collective cavity modes $\hat{A}$ and $\hat{B}$, in the adiabatic limit, in which $\hat{A}^{\dagger}\hat{A} |n \rangle=n_{A}|n \rangle$ and $\hat{B}^{\dagger}\hat{B} |n \rangle=n_{B}|n \rangle$, and $ |\Phi(n,t) \rangle=e^{-i\hat{H}_{\rm{mir}}t/ \hbar}|\Phi(n,0) \rangle$ denotes the time-evolved wave function of the two movable mirrors \cite{cpsun,hian}. The approximation in assigning a system wave function of the form (\ref{mirham1}) lies in the fact that the coefficients $P(n_A,n_B)$ are time-independent, and as a consequence no population transfer occurs between different photon states $|n\rangle$. 

Within this BO approximation, the cavity modes can be seen as inducing an effective potential in which the two mirrors evolve,
\begin{eqnarray}\label{digham2}
\hat{V}_{\rm{eff}} &=& \hbar \left(\omega-g\frac{q_{1}+q_{2}}{2}\right) (n_{A}+n_{B}) \nonumber \\
 &&  -\hbar \sqrt{g^{2}(q_{1}-q_{2})^{2}+4\lambda^{2}}\frac{n_{A}-n_{B}}{2} . 
\end{eqnarray}
Since we have assumed that the oscillation amplitudes of the movable mirrors are small, it follows that their relative displacement  $\hat{q}_{1}-\hat{q}_{2}$ is also small. Therefore, it is sufficient to expand the second term in the cavity Hamiltonian \eqref{digham2} to second (quadratic) order in  $\hat{q}_{1}-\hat{q}_{2}$. This can be justified, since for a typical optomechanical cavity with optical frequency $\omega/2 \pi  \sim 10^{14} ~{\rm Hz}$, length $L \sim 1~{\rm mm}$, mirror frequency $\Omega/ 2 \pi  \sim10^6~{\rm  Hz}$, and with a zero-point-oscillation amplitude of $ 0.02 ~{\rm pm}$, one finds that $g \sim 10^{4} ~{\rm Hz}$. With the reasonable estimate $\lambda=10^{5}~ {\rm Hz}$  one gets $(g/2  \lambda)^{2} \sim 10^{-3}$. This results in an  effective Hamiltonian describing the dynamics of two coupled movable mirror in absence of any losses, 
\begin{eqnarray}\label{mirham2}
\hat{ H}_{\rm{mir}} \approx \hbar \frac{\Omega}{2}\left(\hat{q_{1}}^{2}+\hat{p_{1}}^{2}+\hat{q_{2}}^{2}+\hat{p_{2}}^{2}\right) \\ \nonumber
-(n_{A}-n_{B})\hbar \lambda \frac{g^{2}}{8 \lambda^{2}}(\hat{q_{1}}-\hat{q_{2}})^{2},  \nonumber
\end{eqnarray}
where we have dropped all the constant and classical driving terms from  the Hamiltonian. Dynamical properties of entanglement in a model related to the one of Eq.~(\ref{mirham2}) was recently studied for a closed system~\cite{fernanda}. Equation \eqref{mirham} can be rewritten in terms of
 \begin{eqnarray*}\label{nayaeqn}
\hat{q}_{1}=\frac{(\hat{c}^{\dagger}+\hat{c})}{\sqrt{2} },~~~~~~  \hat{p}_{1}=i\frac{(\hat{c}^{\dagger}-\hat{c})}{\sqrt{2} },\\
\hat{q}_{2}=\frac{(\hat{d}^{\dagger}+\hat{d})}{\sqrt{2} },~~~~~~  \hat{p}_{2}=i\frac{(\hat{d}^{\dagger}-\hat{d})}{\sqrt{2} },\\
\end{eqnarray*}
such that
\begin{eqnarray}\label{mirhamnew}
\hat{ H}_{\rm{mir}}&=& \hbar  \Omega\left(\hat{c}^{\dagger}\hat{c}+\hat{d}^{\dagger}\hat{d}\right) \\
&&-(n_{A}-n_{B}) \hbar  \lambda  \left(\frac{g^{2}}{16 \lambda^{2}}\right)\left(\hat{c}^{2}+\hat{c}^{\dagger 2}+2 \hat{c}^{\dagger} \hat{c}\right) \nonumber \\
&&-(n_{A}-n_{B}) \hbar  \lambda  \left(\frac{g^{2}}{16 \lambda^{2}}\right)\left(\hat{d}^{2}+\hat{d}^{\dagger 2}+2 \hat{d}^{\dagger} \hat{d}\right)\nonumber\\
&&+(n_{A}-n_{B}) \hbar  \lambda  \left(\frac{g^{2}}{8 \lambda^{2}}\right)\left(\hat{c}+\hat{c}^{\dagger}\right)\left(\hat{d}+\hat{d}^{\dagger}\right).\nonumber
\end{eqnarray}
Introducing center-of-mass and relative modes,
\begin{eqnarray}\label{mirhamold}
\hat{C}=\frac{\hat{c}+\hat{d}}{\sqrt{2}},~~~~~~~\hat{D}=\frac{\hat{c}-\hat{d}}{\sqrt{2}},
\end{eqnarray}
Eq.~\eqref{mirhamnew} becomes
\begin{eqnarray}\label{mirhamnewnew}
\hat{ H}_{\rm{mir}} =\hbar  \Omega \hat{C}^{\dagger}\hat{C}+\hbar  ( \Omega-4N \lambda)\hat{D^\dagger}\hat{D} \\ \nonumber  
-2N\hbar \lambda\left(\hat{D}^{2}+\hat{D}^{\dagger2}\right),
\end{eqnarray}
where $N=(n_{A}-n_{B})(g/4 \lambda)^{2}$. The Hamiltonian in  the above form is known to generate  squeezing in the $D$-mode~\cite{sque}, which will also be manifested as quantum correlations between the two mirror oscillations. 

After arriving at this simplified form of the Hamiltonian governing the dynamics of the two movable mirrors, we now provide a fully analytical treatment describing the state evolution of the two mirrors. Firstly we shall discuss the unitary dynamics of the system in section \ref{sec:unitry} and provide a closed-form expression for the time-evolved mirror operators $\hat{c}(t)$ and $\hat{d}(t)$ in the Heisenberg picture. This will allow us to solve for the dynamics of initially uncoupled movable mirrors for an arbitrary initial state. This will then be followed by \ref{sec:disip} where we shall provide  a full solution of the master equation describing the dissipative dynamics of the two indirectly coupled  movable mirrors. 
\subsection{Unitary evolution}
\label{sec:unitry}
The Hamiltonian \eqref{mirhamnewnew} describing the dynamics of the two movable mirrors  can be further diagonalized by a Bogoliubov transformation. We define $\hat{E}$ and $\hat{E}^{\dagger}$ such that
\begin{equation}\label{bogol}
\left(
\begin{array}{c}
\hat{D}^{\dagger} \\
   \hat{D}  
\end{array}
\right)=\left(
\begin{array}{cc}
u& v \\
 v &u 
\end{array}
\right)\left(
\begin{array}{c}
\hat{E}^{\dagger} \\
   \hat{E}  
\end{array}
\right),
\end{equation}
where $u^{2}-v^{2}=1$. Setting 
\begin{eqnarray}\label{eqnuv}
u^{2}=\frac{1}{2}\left(1+\sqrt{1+\frac{4M^{2}}{1-4M^{2}}}\right), \nonumber   \\
v^{2}=\frac{1}{2}\left(-1+\sqrt{1+\frac{4M^{2}}{1-4M^{2}}}\right),
\end{eqnarray}
the Hamiltonian (\ref{mirhamnewnew}) reduces to the 
diagonal form
\begin{equation}\label{eqnuv}
\tilde{H}_{\rm{mir}}= \hbar \Omega \hat{C}^{\dagger} \hat{C}+2 \hbar \omega_{0}\hat{E}^{\dagger} \hat{E},
\end{equation}
where
\begin{eqnarray}\label{eqnuvomeg}
\omega_{0}&=&\frac{(\Omega-8 \lambda N) \Omega}{2 \sqrt{1-4 M^{2}}(\Omega-4\lambda N)}, \nonumber \\
M&=&\frac{2 N\lambda}{\Omega-4 N\lambda}.
\end{eqnarray}
We can then straightforwardly solve the equations of motion for the operators $\hat{C}(t)$ and  $\hat{E}(t)$,
\begin{eqnarray}\label{eqnuv}
\hat{C}(t)&=&\hat{C}(0) e^{i \Omega t}\nonumber\\
\hat{E}(t)&=&\hat{E}(0) e^{-i 2 \omega_{0}t},
\end{eqnarray}
giving the closed expressions for the time evolved operators $\hat{c}(t)$ and $\hat{d}(t)$, 
\begin{eqnarray}\label{eqnuvfinal}
\hat{c}(t) & = & \frac{1}{2}\Big[ F(t)\hat{c}(0)+G(t) \hat{d}(0)
 +2i \sin(2 \omega_0 \it{t})\it{uv}  \hat{\it{c}}^{\dagger}(\rm{0})\nonumber\\
 & &  -{\rm 2}\it{i} \sin({\rm 2} \omega_{\rm 0} \it{t})\it{uv}  \hat{\it{d}}^{\dagger}(\rm{0})\Big] \nonumber\\
\hat{d}(t) & = & \frac{1}{2}\Big[ G(t)\hat{c}(0)+F(t) \hat{d}(0) 
-2i \sin(2 \omega_{0} \it{t})\it{uv}  \hat{\it{c}}^{\dagger}(\rm{0})\nonumber \\
 & & +\rm{2}\it{i} \sin({\rm 2} \omega_{\rm 0} \it{t})\it{uv}  \hat{\it{d}}^{\dagger}(\rm{0})\Big],
\end{eqnarray}
where $F(t)$ and $G(t)$ are  time-dependent complex functions given by 
\begin{eqnarray}\label{eqnuvfinal}
F(t)=e^{-i\Omega t}+ u^{2}e^{-i2\omega_{0}t}-v^{2}e^{i2 \omega_{0}t}, \\ 
G(t)=e^{-i\Omega t}+ v^{2}e^{i2\omega_{0}t}-u^{2}e^{-i2 \omega_{0}t}. 
\end{eqnarray}

With the solution of the operators $\hat{c}(t)$ and $\hat{d}(t)$ now in hand we can faithfully describe the unitary dynamics of the two movable mirrors for any arbitrary initial state.
Of particular interest are  initial  Gaussian states including  thermal, coherent,  and squeezed states.  A Gaussian continuous variable state can be fully described in terms of a real symmetric covariance matrix  {\bf V}.  For a two-mode Gaussian continuous variable system, the covariance matrix {\bf V} can be written as
\begin{eqnarray}
\bf V=
\left(
\begin{array}{cc}
   \bf A  &\bf C   \\
  {\bf C}^{T}  & \bf B   \\
\end{array}
\right),
\end{eqnarray}
where $T$ denotes matrix transpose, 
\begin{eqnarray*}
\bf A\!&=&\!\!\left(\!\begin{array}{cc} \langle (\hat{c}+ \hat{c}^{\dagger})^2\rangle /2& \langle [\hat{c}+\hat{c}^{\dagger},i(\hat{c}^{\dagger}-\hat{c})]_{+}\rangle/2 \\   \langle [\hat{c}+\hat{c}^{\dagger},i(\hat{c}^{\dagger}-\hat{c})]_{+}\rangle/2 & \langle (i(\hat{c}^{\dagger}-\hat{c}))^2\rangle /2 \end{array}\right)\!, \\ \\ \nonumber
\bf B\!&=&\!\!\left(\!\begin{array}{cc} \langle (\hat{d}+ \hat{d}^{\dagger})^2\rangle /2& \langle [\hat{d}+\hat{d}^{\dagger},i(\hat{d}^{\dagger}-\hat{d})]_{+}\rangle/2 \\  \langle [\hat{d}+\hat{d}^{\dagger},i(\hat{d}^{\dagger}-\hat{d})]_{+}\rangle/2 & \langle (i(\hat{d}^{\dagger}-\hat{d}))^2\rangle /2 \end{array}\!\right)\!, \\ \\ \nonumber
\bf C\!&=&\!\!\left(\!\begin{array}{cc} \langle (\hat{c}+ \hat{c}^{\dagger})(\hat{d}+ \hat{d}^{\dagger})\rangle /2& \langle i(\hat{c}+ \hat{c}^{\dagger})(\hat{d}^{\dagger}-\hat{d}) \rangle/2 \\  \langle(i(\hat{c}^{\dagger}-\hat{c}) (\hat{d}+ \hat{d}^{\dagger})\rangle /2 & \langle- (\hat{c}^{\dagger}-\hat{c})(\hat{d}^{\dagger}-\hat{d})\rangle /2 \end{array}\right)\!, \\ \nonumber
\end{eqnarray*}
and $\langle [\hat{r}_{i},\hat{r}_{j}]_{+} \rangle= (\langle \hat{r}_{i}\hat{r}_{j}+\hat{r}_{j}\hat{r}_{i}) \rangle)/2$. Once we have the covariance matrix, all the quantum statistical properties of Gaussian continuous variable states can be constructed. Also worth  mentioning is the important fact that the Hamiltonian \eqref{mirhamnew} is quadratic in the position and momentum coordinates of the movable mirrors. An initial Gaussian state of the mirror evolving under \eqref{mirhamnew} will therefore maintain its Gaussian character. 

\begin{figure}[!]
\centering
\includegraphics[width=0.451\textwidth]{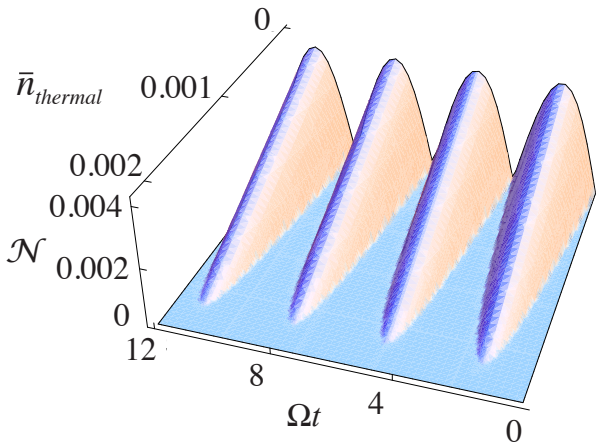}
\caption{\label{fig1}(Color online) Time evolution of the  degree of entanglement, as measured by the logarithmic  negativity, as a function of initial temperature of the movable mirrors, measured in terms of $\bar{n}_{\rm thermal}$.  The dimensionless parameters are chosen such that $\Omega=1$, $g=10^{-2}$, $\lambda=10^{-1}$, $\alpha_{A}=4$ and $\alpha_{B}=1$.}
\end{figure}

A widely used entanglement measure is the logarithmic negativity, which is an entanglement monotone and fairly  easy to compute \cite{gerado}. For a two-mode Gaussian continuous variable state characterized by the  covariance matrix $\bf V$, logarithmic negativity is defined as
\begin{equation}\label{negtyeqn}
 \mathcal N= \rm{Max}[0,log(2 \tilde{\nu}_{-})],
\end{equation}
where $\tilde{\nu}_{-}$ is the smallest of the symplectic eigenvalues of the partially transposed covariance matrix given by  
\begin{equation}\label{negty}
\begin{array}{l}
 \tilde{\nu}_{-}= \sqrt{\sigma/2-\sqrt{(\sigma^{2}-4 \rm{Det} \bf{V})/\rm{2}}} \\ \\
\sigma=\rm{Det} \bf{A}+\rm{Det} \bf{B}-\rm{2Det} \bf{C}. 
\end{array}
\end{equation}
We analytically reconstruct the time-dependent covariance matrix, from which  it is then straightforward to compute the logarithmic negativity, with a typical solution shown in Fig.~\ref{fig1}. In these calculations, the logarithmic negativity has been weighted  with a coherent state probability distribution for the collective cavity modes $A$ and $B$ such that $P(n_{A},n_{B})={\rm exp}(-(|\alpha_{A}|^{2}+|\alpha_{B}|^{2})) |\alpha_{A}| ^{{\rm 2}n_{A}}|\alpha_{B} |^{{\rm 2}n_{B}}/(n_{A}! n_{B}!)$. Such averaging accounts for initial quantum fluctuations in the two cavity modes. A non-zero value of $\mathcal N$ quantifies the degree of entanglement between the two movable mirrors.  As can be seen from Fig.~\ref{fig1}, increasing the initial temperature of the mirror degrades the quantum correlations and eventually leads to completely separable states of the two mirrors. The figure also gives a clear example of entanglement sudden death and birth~\cite{esd}, arising from the common coupling of the mirrors to the two cavity modes.

\subsection{Dissipative dynamics}
\label {sec:disip}
In any physical setting, coupling to the environment is inevitable and typically results in decoherence of the quantum state to its classical counterpart. In the scheme of interest to us, there can be two main causes of dissipation. One is the photon leakage through the two cavities, and the other is
thermal decay of the states of the movable mirrors due to their coupling to baths of non-zero temperature. 

So far we have treated the cavity modes in the adiabatic approximation by expanding the global wave function in terms of energy eigenstates of the collective cavity operators $\hat{A}$ and $ \hat{B}$. Although it might look somewhat artificial at first, this approach has its own advantages. Firstly, it allows us to derive a closed-form analytical result governing the dynamics of the two coupled mirrors, which is otherwise a nonlinear problem in itself. Secondly, it also provides us with useful physical insights into how the cavity assisted entanglement of spatially separated mechanical oscillators originates. In the present section we will continue treating the two cavity modes in this semiclassical regime and defer the explicit calculations involving reservoir induced quantum fluctuations in the cavity modes till the next section. More precisely, we assign coherent states for the two cavity modes and introduce cavity losses  in terms of a non-Hermitian Hamiltonian.

A phenomenological way to introduce cavity losses is to shift the cavity resonance frequency $\omega$ by $-i \kappa$ where $\kappa$ is the cavity decay rate. Then, under the BO approximation, the two indirectly coupled movable mirrors evolve according to the non-Hermitian Hamiltonian
\begin{equation}\label{mirrodispnaya}
\hat{H}_{\rm disp} \approx \hat{H}_{\rm mir}-i \kappa (n_{A}+n_{B})\,,
\end{equation}
where $\hat{H}_{\rm mir}$ is given by \eqref{mirhamnew} and we again have neglected all the classical driving terms in the Hamiltonian. Apart from the cavity losses, the two cavity mirrors might  undergo further decoherence due to their inevitable coupling to the external environment. The time evolution of the mixed state of the two movable mirrors obtained by tracing over the cavity field  distribution takes the form
\begin{equation}\label{timevldensmat}
\hat{\rho}_{\rm mir}(t)=\!\frac{1}{\sum_{n_{A}, n_{B}} N_{n_A,n_B}(t)}\sum_{n_{A}, n_{B}} N_{n_A,n_B}(t) \hat{\rho}_{\rm mir}^{(n)}(t),
\end{equation}
where
\begin{eqnarray}
N_{n_A,n_B}(t) & = & \exp\left(-|\alpha_{A}e^{-\kappa t }|^{2}\right)|\alpha_{A}e^{-\kappa t }| ^{{\rm 2}n_{A}} \\
& & \times\exp\left(-|\alpha_{B}e^{-\kappa t }|^{2}\right)|\alpha_{B}e^{-\kappa t }| ^{{\rm 2}n_{B}}/n_{A}!n_{B}!,\nonumber
\end{eqnarray}
and $\hat{\rho}_{\rm mir}^{(n)}(t)$ is the time-evolved reduced density matrix of the two movable mirrors with the photon number difference $n=n_A-n_B$. It turns out that if both collective cavity modes are initially coherent states, this particular method of taking dissipation into account is not only accurate but also exact \cite{qjump}. This is because an initial coherent state evolving in a purely dissipative channel remains a coherent state, although with an exponentially decaying amplitude \cite{coherent}. The time evolution of $\hat{\rho}_{\rm mir}^{(n)}(t)$ in the  Born-Markov approximation is then described by the Lindblad-type master equation \cite{mil,stev}
\begin{eqnarray}\label{masterpurana}
\frac{\partial}{\partial t}\hat{\rho}_{\rm mir}& =& -i\left[\hat{H}_{\rm mir},\hat{\rho}_{\rm mir}\right] +\frac{\Gamma}{2}\bar n\mathcal L_{c^\dagger}\hat{\rho}_{\rm mir}+\frac{\Gamma}{2} \bar n \mathcal L_{d^\dagger}\hat{\rho}_{\rm mir} \nonumber\\
& & +\frac{\Gamma}{2}(\bar n+1)\mathcal L_{c}\hat{\rho}_{\rm mir} +\frac{\Gamma}{2}(\bar n+1)\mathcal L_{d}\hat{\rho}_{\rm mir},
\end{eqnarray}
where $\Gamma$ is the decay rate of each movable mirror due to its coupling to a heat 
bath with average thermal occupancy  $\bar n$, and
$\mathcal L_{x}\hat{\rho} \equiv 
2\hat{x}\hat{\rho} \hat{x}^{\dagger}-\hat{x}^{\dagger}\hat{x}\hat{\rho}-\hat{\rho} \hat{x}^{\dagger} \hat{x}$.
In terms of the center-of-mass mode $\hat{C}$ and relative mode  $\hat{D}$, Eq.~\eqref{masterpurana} can be equivalently written as
\begin{eqnarray}\label{masternew}
\frac{\partial \rho_{\rm mir} }{\partial t} & = & -i\left[\hat{H}_{\rm mir},\hat{\rho}_{\rm mir}\right]+\frac{\Gamma}{2} \bar n\mathcal L_{C^\dagger}\hat{\rho}_{\rm mir}+\frac{\Gamma}{2} \bar n\mathcal L_{D^\dagger}\hat{\rho}_{\rm mir} \nonumber\\ 
& & +\frac{\Gamma}{2}(\bar n+1)\mathcal L_{C}\hat{\rho}_{\rm mir}+\frac{\Gamma}{2}(\bar n+1)\mathcal L_{D}\hat{\rho}_{\rm mir},
\end{eqnarray}
where $\hat{H}_{\rm mir}$ is given by \eqref{mirhamnewnew}.

To solve the master equation \eqref{masternew}, we define the normal-ordered quantum characteristic function \cite{mil,stev} for the two movable mirrors as $
\chi(\epsilon,\eta,t)=\langle e^{\epsilon \hat{C}^{\dagger} }e^{-\epsilon ^{*}\hat{C}} e^{\eta \hat{D}^{\dagger}} e^{-\eta^{*} \hat{D}}\rangle$.
Using standard quantum optical techniques \cite{mil,stev}, the master equation \eqref{masternew} can be rewritten as a partial differential equation for the quantum characteristic function $\chi(\epsilon,\eta,t)$ of the form
\begin{equation}\label{chievnay}
\frac{\partial }{\partial t}\chi(\epsilon,\eta,t)
=z^{T}  {\mbox{\bf M}} \nabla \chi(\epsilon,\eta,t)+\rm{4} \lambda \it{z}^{T}  {\mbox{\bf K}} \it{z} \chi(\epsilon,\eta,t),
\end{equation}
where 
\begin{equation}
\begin{array}{l}
z^{T}=(u_{1},u_{2},v_{1},v_{2}), \\ \\
\nabla= \left(\frac{\partial}{ \partial u_{1}}, \frac{\partial }{ \partial u_{2}}, \frac{\partial}{\partial v_{1}}, \frac{\partial }{ \partial v_{2}}\right)^{T} , \\ \\
u_{1}=\frac{\epsilon_{c}+\epsilon_{d}+\epsilon_{c}^{*}+\epsilon_{d}^{*}}{2\sqrt{2}}, \\ \\  u_{2}=\frac{\epsilon_{c}+\epsilon_{d}-\epsilon_{c}^{*}-\epsilon_{d}^{*}}{i2\sqrt{2}}, \\ \\ 
v_{1}=\frac{\epsilon_{c}-\epsilon_{d}+\epsilon_{c}^{*}-\epsilon_{d}^{*}}{2\sqrt{2}},\\ \\  v_{2}=\frac{\epsilon_{c}-\epsilon_{d}-\epsilon_{c}^{*}+\epsilon_{d}^{*}}{i2\sqrt{2}}, 
\end{array}
\end{equation}
and
\begin{eqnarray}
\epsilon_{c}=\frac{\epsilon+\eta}{\sqrt{2}}, & \hspace{1cm} & \epsilon_{d}=\frac{\epsilon-\eta}{\sqrt{2}}.
\end{eqnarray}
The $4\times4$ matrix coefficients of Eq.~(\ref{chievnay}) read
\begin{equation}
\begin{array}{llll}
{\mbox{\bf M}}=\left(\begin{array}{cc}
    {\mbox{\bf M}_{1}} & {\mbox{\bf 0}} \\ 
   {\mbox{\bf 0}} &   {\mbox{\bf M}_{2}}  \\ 
  \end{array}\right), & & &
  {\mbox{\bf K}}=\left( \begin{array}{cc}
    {\mbox{\bf K}_{1}} & {\mbox{\bf 0}} \\ 
   {\mbox{\bf 0}} &   {\mbox{\bf K}_{2}}  \\ 
  \end{array} \right ),
\end{array}
\end{equation}
with
\begin{equation}
\begin{array}{l}
{\mbox{\bf M}_{1}}=\left(\begin{array}{cc}
    {-\Gamma/2} & {\Omega} \\ 
   {-\Omega} &   {-\Gamma/2}  \\ 
  \end{array}\right),\\ \\
    {\mbox{\bf M}_{2}}=\left( \begin{array}{cc}
    {-\Gamma/2} & {\Omega-8 N \lambda} \\ 
   {-\Omega} &   {-\Gamma/2}  \\ 
  \end{array} \right ),\\ \\
  {\mbox{\bf K}_{1}}=\left( \begin{array}{cc}
    {-\Gamma \bar n/ 4 \lambda} & {0} \\ 
    {0} &{-\Gamma \bar n/ 4 \lambda}  \\ 
  \end{array} \right ),\\ \\
    {\mbox{\bf K}_{2}}=\left( \begin{array}{cc}
    {-\Gamma \bar n/ 4 \lambda} & {N} \\ 
    {N} &{-\Gamma \bar n/ 4 \lambda}  \\ 
  \end{array} \right ).
\end{array}
\end{equation}
For an initial Gaussian state of the two movable mirrors, it is consistent to make the following Ansatz for the quantum characteristic function,
\begin{equation}
\chi(\epsilon,\eta,t)=\exp\left[-z^{T}{\mbox{\bf L}}(t)\,z+i z^{T}q(t)\right],
\end{equation}
where ${\mbox{\bf L}}(t)$ is a $4 \times 4 $ time-dependent symmetric matrix and $q(t)$ is a $4 \times 1$ time-dependent vector. Using the above Ansatz in Eq.~\eqref{chievnay} results in the coupled matrix differential equations
   \begin{eqnarray}
  \dot{{\mbox{\bf L}}}&=&{\mbox{\bf M}} {\mbox{\bf L}}+{\mbox{\bf L}}{\mbox{\bf M}}^{T}-4 \lambda {\mbox{\bf K}}, \\ 
 \dot{q}&=&{\mbox{\bf M}}q.
  \end{eqnarray}
  Making use of the fact that ${\mbox{\bf L}}$ is a  $4 \times 4$ symmetric matrix, it can be decomposed into $2 \times 2$ square matrices such that
  \begin{eqnarray}
 {\mbox{\bf L}}=\left( \begin{array}{cc}
    { {\mbox{\bf P}}} & { {\mbox{\bf Q}}} \\ 
    { {\mbox{\bf Q}^{T}}} &{ {\mbox{\bf R}}}  \\ 
  \end{array} \right ),
\end{eqnarray}
where ${\mbox{\bf P}}$ and ${\mbox{\bf R}}$ are 2$\times$2 symmetric matrices. Obtaining an explicit form for the time-dependent  quantum characteristic function $\chi(\epsilon,\eta,t)$ now reduces to solving $2 \times 2$ coupled matrix differential equations. 

Although an exact analytical solution can be arrived at, it is too lengthy  to be reported here. Nonetheless, the time-dependent covariance matrix $ {\mbox{\bf V}}$ can be fully reconstructed from the quantum characteristic function $\chi(\epsilon,\eta,t)$. This can be  easily seen by noting that  from the quantum characteristic function one can obtain the expectation values of quantum mechanical observables, e.g., 
 $\langle \hat{c}^{\dagger m}(t){\it \hat{d}^{\dagger n}}(t) \rangle=\it {(\frac{\partial }{\partial \epsilon_{c}})^{m}(\frac{\partial }{\partial \epsilon_{d}})^{n}\chi(\epsilon_{c},\epsilon_{d},t)}|_{\it{\epsilon_{c},\epsilon_{d}}=\rm{0}}$ and thus all the elements of the  covariance matrix can be found. 
 
\begin{figure}[!]
\centering
\includegraphics[width=0.451\textwidth]{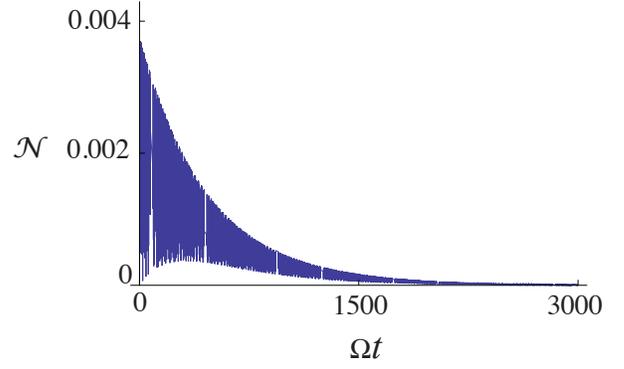}
\caption{\label{fig2}(Color online) Temporal evolution of the degree of entanglement between two indirectly coupled movable mirrors as measured by the logarithmic  negativity.
Compared with Fig.~\ref{fig1}, here losses in  all modes have been considered and the degree of entanglement is consequently somewhat smaller, but importantly, it survives for a reasonably  long time.  Each mirror is initially assumed to be in its ground state and the dimensionless parameters used are chosen such that $\Omega=1$, $g=10^{-2}$, $\lambda=10^{-1}$, $\alpha_{A}=4$, $\alpha_{B}=1$, $\kappa=10^{-3}$, $\Gamma=10^{-4}$ and $\bar{n}=0$.}\end{figure}

As a measure of entanglement between the distant cavity mirrors we again compute the logarithmic negativity. 
The result of such a calculation is shown in Fig.~\ref{fig2}. As is clear from the figure, under the action of cavity-mediated coupling, the two movable mirrors exhibit entanglement. Although the entanglement generated is not too large, 
it is sustained over a reasonably long timescale. The degree of inseparability  between the two  mirrors can be improved significantly either by a conditional measurement of the cavity field or by increasing the difference in the mean number of photons in the  field distributions of the two cavity modes. Thus we conclude that the aforementioned protocol is indeed capable of generating quantum entangled states of two movable mirrors, which are robust with respect to dissipation 
for a long time. It should be pointed out that the logarithmic negativity approaches zero exponentially for large times due to the decay of photons out of the cavities. In order to have sustainable non-vanishing entanglement, the photon modes must be driven externally to prevent the absence of photons. This will be discussed in the following section. 

\subsection{Validity of the Born-Oppenheimer approximation}
\label{sec:BOvalidity}

So far we have used the BO approximation to separate the slow dynamics of the two movable mirrors from the rapidly evolving population of the two cavity modes. The BO approach has proved to be a fundamental tool in various applications of physics and quantum chemistry \cite{BO,lsced}. It is nothing but an extension of the quantum adiabatic theorem to a quantum system with two sets of variables whose dynamics can be separated due to very different dynamical  time scales.

The original version of the adiabatic theorem asserts that an initial eigenstate of a slowly varying Hamiltonian will remain in the same instantaneous eigenstate throughout the evolution. In our system, the absence of transitions between instantaneous eigenstates translates to fixed probabilities for the states $|n\rangle$, see Eq.~(\ref{mirham1}), to be occupied. There has been continued interest in the application of the adiabatic theorem in slowly evolving quantum mechanical systems including Berry's phase and effective gauge theories~\cite{berry,gauge}, geometric quantum computation~\cite{geom}, and adiabatic quantum computation~\cite{adiaba}.  However, recently the generally accepted criteria for adiabaticity in quantum physics has suffered some criticisms~\cite{criti}, which  has lead to various arguments and 
counter-arguments to ascertain the justification of the adiabatic approximation in quantum theory~\cite{dmtong, mhsamin}. A breakdown of adiabaticity seems to appear only in rather special cases of rapid resonant driving of the system~\cite{mhsamin}. This kind of external resonant driving is absent in our model and we expect a standard
application of the adiabatic theorem, or more precisely an application of the BO approximation, to be unproblematic. In the beginning of Sec.~\ref{ssec:admod}, we pointed out that the two types of oscillators evolve on very different time scales which normally guarantees adiabatic dynamics. However, one issue overlooked so far concerns the fact that we consider an open quantum system for which the concept of adiabaticity must be handled with care~\cite{xlhunag}. Loosely speaking, one   could imagine non-adiabatic transitions induced by the reservoir. In the following discussion we will argue that such transitions are indeed not present in our model. 

Recently an extension of the BO approximation has been presented for a quantum system  coupled to a large reservoir \cite{xlhunag}. Following this work, we can write the Hamiltonian  of the closed quantum system as $H=H_{s}(X)+H_{f}(X,Y)$, where $H_{s}(X)$ is the Hamiltonian of the slowly varying dynamical variable $X$ and $H_{f}(X,Y)$ is the interaction Hamiltonian between the slowly varying variable and the quickly varying dynamical variable $Y$. If the coupling between the quantum  system  and the reservoir is such that its evolution can be described using the Lindblad approach, then, in the case of  no quantum jumps, the evolution of the quantum system  is governed by the equation
\begin{equation}
i \hbar \frac{d}{dt} |\Psi(t) \rangle=H_{\rm eff}|\Psi(t) \rangle.
\end{equation}
Here $H_{\rm eff}=H_{s}(X)+H_{f}'(X,Y)$, and $H_{f}'(X,Y)$ is the non-hermitian part of the Hamiltonian, with dissipation assumed to affect only 
the quickly varying variable of the Hamiltonian. Treating the slowly varying variable $X$ as a parameter, we can solve for left and 
right eigenstates $\langle \Psi_{n,X}^{L}(Y) |$ and $|\Psi_{n,X}^{R}(Y) \rangle$ of the non-hermitian Hamiltonian $H_{f}'(X,Y)$, with complex eigenvalues $E_{n,X}(Y)$. 
Expanding the eigenstates of $H_{\rm eff}$ in terms of $|\Psi_{n,X}^{R}(Y) \rangle$ 
we get 
\begin{equation}
|\Phi \rangle= \sum_{n=1}^{N} c_{n}|\varphi_{n}(X)\rangle |\Psi_{n,X}^{R}(Y) \rangle,
\end{equation}
where $N$ is the dimension of the quickly varying variable $Y$ and $c_{n}$ are the expansion coefficients. It can be shown that the eigenvalue equation $ H_{\rm eff}|\Phi \rangle=E |\Phi \rangle$ can be recast in the form
$(H_{o}+H_{P})\varphi = E \varphi $ \cite{xlhunag}, where
\begin{eqnarray*}
H_{o}(X)
\!&=&\!\! 
\left[ \begin{array}{cccc} 
\!H_{1}\!
+\!
E_{1}(Y) & 0&\cdots&0\\
 0 &H_{2}\!
 +\!
 E_{2}(Y) &\cdots&0\\
\vdots&\vdots&\ddots&\vdots \\
0 &0 & \cdots & H_{N}\!
+\!
E_{N}(Y)\! 
\end{array}
\right]\!,
 \\
H_{P}(X) &=& \left[ \begin{array}{cccc} 
0& H_{1,2}&\cdots&H_{1,N}\\
 H_{2,1}&0&\cdots&H_{2,N}\\
\vdots&\vdots&\ddots&\vdots \\
H_{N,1}&H_{N,2}&\cdots&0
\end{array}
\right] , 
~~\varphi=  \left[ \begin{array}{c} 
|\varphi_{1} \rangle\\
|\varphi_{2} \rangle\\
\vdots \\
|\varphi_{N} \rangle
\end{array}
\right]
\end{eqnarray*}
Here $H_{n}=H_{n,n}$, $E_{n}(Y)=E_{n,X}(Y)$ and $H_{n,m}(X)=\langle \Psi_{n,X}^{L}(Y) | H_{s}(X) |\Psi_{n,X}^{R}(Y) \rangle$. Treating $H_{P}$ as a small perturbation  and using 
standard time independent perturbation theory, corrections to all orders   for the eigenstates and the eigenenergies can be calculated. Since there is no resonant driving, we can conclude that the BO approximation is accurate as long as the  condition 
  \begin{equation}
  \left|\frac{\langle \Phi_{n',k'}^{L[0]} | H_{n',n}|\Phi_{n,k}^{R[0]} \rangle}{E_{n',k'}^{[0]}-E_{n,k}^{[0]}}\right| << 1 ~~~~~~~{\rm for~all~}~k', n' \neq k,n.
  \end{equation}
is satisfied \cite{xlhunag}.

Returning to consider our original  physical system of two coupled optomechanical cavities, the non-hermitian Hamiltonian containing the cavity decay terms takes the 
form
\begin{eqnarray}\label{crct}
\hat{H}_{\rm{cav}} & = &  
{\hbar \left (\omega-i \kappa-g\frac{\hat{q}_{1}+\hat{q}_{2}}{2}\right) (\hat{A}^{\dagger}\hat{A}+\hat{B}^{\dagger}\hat{B})} \nonumber \\ 
  && 
  {- \hbar \sqrt{g^{2}{\left (\hat{q}_{1}-\hat{q}_{2}\right)^{2}+4\lambda^{2}}} \frac{(\hat{A}^{\dagger}\hat{A}-\hat{B}^{\dagger}\hat{B})}{2}}  \nonumber \\
 && 
 {+\hbar \frac{\Omega}{2} (\hat{q}_{1}^{2}+\hat{p}_{1}^{2}+\hat{q}_{2}^{2}+\hat{p}_{2}^{2})}. 
\end{eqnarray}
The quickly varying part of the Hamiltonian 
\begin{eqnarray*}
H_{f}&=&\hbar \left[\omega-i \kappa-g(\hat{q}_{1}+\hat{q}_{2})/2\right] (\hat{A}^{\dagger}\hat{A}+\hat{B}^{\dagger}\hat{B})- \\
&&\hbar \sqrt{g^{2}{\left (\hat{q}_{1}-\hat{q}_{2}\right)^{2}+4\lambda^{2}}} (\hat{A}^{\dagger}\hat{A}-\hat{B}^{\dagger}\hat{B})/2
\end{eqnarray*}
has 
left  and 
right eigenstates 
$\langle \Psi_{n,X}^{L}(Y) |=\langle n_{A} | \langle n_{B} | $  and $|\Psi_{n,X}^{R}(Y) \rangle=|n_{A} \rangle |n_{B} \rangle$ respectively, 
 with complex eigenvalues given by 
\begin{eqnarray*}
E_{n,X}(Y)&=&\hbar (\omega-i \kappa-g (\hat{q}_{1}+\hat{q}_{2})/2) (n_{A}+n_{B})- \\
&&\hbar \sqrt{g^{2}(\hat{q}_{1}-\hat{q}_{2})^{2}+4\lambda^{2}}(n_{A}-n_{B})/2.
\end{eqnarray*}
  As a passing remark, 
the perturbation terms $H_{P}$ are in general non-zero. This demands that the dissipation in the slowly varying variables must be slower than 
the dissipation in the quickly varying variables. If this condition is not met, then  
strong dissipation in the slowly varying variables enlarges the perturbation $H_{P}$, which might 
make the BO approximation invalid. Roughly 
speaking, a stronger rate of dissipation in the slowly varying variables accelerates their rate of change,
thus making it hard to distinguish their dynamics  from that of the quickly varying variables.
The slowly varying variables  then evolve under the 
 effective Hamiltonian
  \begin{eqnarray}\label{crctnew}
\hat{H}_{\rm mirror} & =& \hbar \frac{\Omega}{2}\left(\hat{q}_{1}^{2}+\hat{p}_{1}^{2}+\hat{q}_{2}^{2}+\hat{p}_{2}^{2}\right) \nonumber \\
&& - \hbar \sqrt{g^{2}{\left (\hat{q}_{1}-\hat{q}_{2}\right)^{2}+4\lambda^{2}}} \frac{(n_{A}-n_{B})}{2}  \nonumber \\
&&  +\hbar \left (\omega-i \kappa-g\frac{\hat{q}_{1}+\hat{q}_{2}}{2}\right) (n_{A}+n_{B}).
\end{eqnarray}
It should
be noticed 
that in this 
treatment, the off-diagonal terms of the perturbation $H_{P}$,  which are given by $H_{n,m}(X)=\langle n_{A} | \langle n_{B}| \hat{H}_{\rm mirror }| m_{A}  \rangle | m_{B}\rangle$, are identically zero. This implies that the zeroth order BO approximation for this particular physical model is not only accurate but 
also exact. The conclusion remains the same even  when 
dissipation in the slow variables is taken into account \cite{xlhunag}.

\section{Quantum Langevin approach}
\label{sec:langv}
As mentioned before, in a typical optomechanical setup, due to coupling induced by radiation pressure, a movable mirror interacts with a cavity mode.   
Unfortunately, the radiation pressure coupling for an undriven cavity with a movable mirror is usually  very weak. This problem can be circumvented by driving the cavity with a coherent classical laser field. Driving
with an intense laser field enhances the radiation pressure coupling and thus facilitates the observation of non-classical phenomena such as entanglement between mechanical oscillators and light. 

In the previous section, cavity driving was not taken into account. Instead the two cavity fields were assumed to be initially in coherent states and the system time-evolution in the presence of the decay of the two cavity fields was studied. In what follows, we shall instead study the situation of cavity driving and show that robust steady state entanglement may exist between different optical and mechanical modes. To this end, we find it more convenient to work in the Heisenberg picture. 
For our system of two coupled cavities with movable mirrors, the Hamiltonian in the interaction picture of the two driving lasers with  frequency $\omega_{L}$ now takes the  
form
\begin{eqnarray}
\label{linham}
\frac{\hat{H}}{\hbar} &=& (\omega-\omega_{L})(\hat{a}^{\dagger}\hat{a}+\hat{b}^{\dagger}\hat{b}) 
 +\frac{\Omega}{2}(\hat q_{1}^{2}+\hat{p_{1}}^{2})+\frac{\Omega}{2}(\hat q_{2}^{2}+\hat p_{2}^{2}) \nonumber\\
&&+\lambda(\hat{a}^{\dagger}\hat{b}+\hat{b}^{\dagger}\hat{a})
-g\hat{a}^{\dagger}\hat{a}\hat{q_{1}}-g\hat{b}^{\dagger}\hat{b}\hat{q_{2}} \nonumber\\
 &&+i \eta(\hat{a}^{\dagger}-\hat{a})+i \eta(\hat{b}^{\dagger}-\hat{b}).
\end{eqnarray}
Here $\eta=\sqrt{2P_{c}\kappa/\hbar \omega_{c}}$ is related to the  driving laser, where $P_{c}$ is the power of the driving laser and $\kappa$ is the damping rate, identical for both cavities.

The Hamiltonian~\eqref{linham} describes the closed-system dynamics of the two driven coupled cavities with movable mirrors. However, as discussed in the previous section, the dynamics of the system is also affected by damping and noise. The main channels of dissipation in our system are the decay in the cavity modes and the coupling of the movable mirrors to their independent thermal baths~\cite{combath}.  One possible way to take into account all the damping and noise processes is to use quantum Langevin equations (QLEs). The QLEs  are equivalent to the Heisenberg equations of motion for time-evolving operators, where noise and dissipative processes have been included phenomenologically~\cite{mil}. 

For the Hamiltonian \eqref{linham}, the QLEs for the cavity and the mirror modes  become
\begin{equation}\label{heilan}
\begin{array}{l}
\frac{d \hat{a}}{d t}=-i \lambda \hat{b}+i g \hat{q}_{1} \hat{a}+\eta-(\kappa+i \tilde{\Delta})\hat{a}+\sqrt{2 \kappa} \hat{a}_{in},\\ \\ 
\frac{d \hat{b}}{d t}=-i \lambda \hat{a}+i g \hat{q}_{2} \hat{b}+\eta-(\kappa+i \tilde{\Delta})\hat{b}+\sqrt{2 \kappa} \hat{b}_{in}, \\ \\
\frac{d \hat{p}_{1}}{d t}=-\Omega \hat{q}_{1}+g \hat{a}^{\dagger} \hat{a}-\gamma_{m}\hat{p}_{1} +\hat{\varepsilon}_{1}(t),\\ \\
\frac{d \hat{p}_{2}}{d t}=-\Omega \hat{q}_{2}+g \hat{b}^{\dagger} \hat{b}-\gamma_{m}\hat{p}_{2} +\hat{\varepsilon}_{2}(t),\\ \\
\frac{d \hat{q}_{1}}{d t}=\Omega \hat{p}_{1}, \\ \\
\frac{d \hat{q}_{2}}{d t}=\Omega \hat{p}_{2}, 
\end{array}
\end{equation}
where $\tilde{\Delta}=\omega-\omega_{L}$ is the  laser detuning from the  cavity resonance frequency $\omega$,   $\kappa$  is the decay rate of each cavity and   $\gamma_{m}$ is the thermal decay rate, identical for the two movable mirrors 
 subjected to independent Brownian motion noise characterized by the operators  $\hat{\varepsilon}_{1}(t)$ and $\hat{\varepsilon}_{2}(t)$ respectively. The quantum noise operators have the 
quantum statistical properties
\begin{eqnarray}\label{brownina}
\langle \hat{\varepsilon}_{1}(t) \rangle&=&
\langle \hat{\varepsilon}_{2}(t) \rangle=0,\\
\langle \hat{\varepsilon}_{i}(t) \hat{\varepsilon}_{j}(t') \rangle & = & \frac{\gamma_{m}}{\Omega}\int e^{-i \omega^{'}(t-t^{'})}\omega^{'}\nonumber\\
& & \times\left[1+\coth\left(\frac{\hbar \omega^{'}}{k_{B}T_{i}}\right)\right]\frac{d \omega^{'}}{2 \pi}\delta_{ij},\nonumber
\end{eqnarray}
where $i,\,j\in 1,\,2$ and $\delta_{ij}$ is the Kronecker delta. We have also introduced independent cavity  input noise operators, $\hat{a}_{in}(t)$ and $\hat{b}_{in}(t)$ for the first and second cavity  respectively. For the case of optical fields, $\hbar \omega/ k_{B} T \gg1$, and hence the mean number of thermal photons can be safely neglected. In this limit the noise operators $\hat{a}_{in}(t)$ and $\hat{b}_{in}(t)$ satisfy the two-time correlations 
\begin{equation}\label{cavtynise}
\begin{array}{l}
\langle \hat{a}_{in}(t)\hat{a}^{\dagger}_{in}(t') \rangle=\delta(t-t'),\\ \\
\langle \hat{a}^{\dagger}_{in}(t)\hat{a}_{in}(t') \rangle=0,\\ \\
\langle \hat{b}_{in}(t)\hat{b}^{\dagger}_{in}(t') \rangle=\delta(t-t'),\\ \\
\langle \hat{b}^{\dagger}_{in}(t)\hat{b}_{in}(t') \rangle=0.
\end{array}
\end{equation}
We are interested in investigating the possibility of achieving steady-state entanglement between distant optical and  mechanical modes. To pursue this aim we have to solve the set of coupled nonlinear  QLEs \eqref{heilan}. This task is difficult, but it is simplified in the presence of strong external driving, in which case linearization of the set of QLEs around the steady-state values is justified. Solving the set of QLEs \eqref{heilan} for the steady-state amplitudes of the optical and mechanical modes, we get
\begin{equation}\label{stdystst}
\begin{array}{l}
q_{1}^{s}=\frac{g|a_{s}|^{2}{\Omega}}, \\ \\
q_{2}^{s}=\frac{g|b_{s}|^{2}{\Omega}}, \\ \\
p_{1}^{s}=0, \\ \\
p_{2}^{s}=0, \\ \\
{a}_{s}=\frac{-i \lambda \eta +\eta(\kappa+i \Delta_{b})}{\lambda^{2}+\kappa^{2}+i \kappa(\Delta_{a}+\Delta_{b})-\Delta_{a} \Delta_{b}}, \\ \\
{b}_{s}= \frac{-i \lambda \eta +\eta(\kappa+i \Delta_{a})}{\lambda^{2}+\kappa^{2}+i \kappa(\Delta_{a}+\Delta_{b})-\Delta_{a} \Delta_{b}},\\ \\
\Delta_{a}=\tilde{\Delta}-g q_{1}^{s}=\omega-\omega_{L}-g q_{1}^{s}, \\ \\
\Delta_{b}=\tilde{\Delta}-g q_{2}^{s}=\omega-\omega_{L}-g q_{2}^{s}.
\end{array}
\end{equation}
In the regime where the two cavities are very intensely driven, such that $|a_{s}|, |b_{s}| \gg 1$, and by expanding the mechanical and optical mode operators as quantum fluctuations around the steady state values ($\hat{a}=a_{s}+\delta \hat{a}$, $\hat{b}=b_{s}+\delta \hat{b}$, $\hat{q}_{i}=q_{i}^{s}+\delta \hat{q}_{i}$ and $\hat{p}_{i}=p_{i}^{s}+\delta \hat{p}_{i}$ for $i=1,2$) we obtain the following linearized QLEs for the quantum fluctuations,
\begin{equation}\label{lineareq}
\begin{array}{l}
\frac{d \delta\hat{a} }{d t}=-\delta \hat{a}(\kappa+i \Delta_{a})-i \lambda \delta \hat{b}+i g a_{s} \delta \hat{x}+\sqrt{2 \kappa} \hat{a}_{in},\\ \\
\frac{d \delta\hat{b} }{d t}=-\delta \hat{b}(\kappa+i \Delta_{b})-i \lambda \delta \hat{a}+i g b_{s} \delta \hat{y}+\sqrt{2 \kappa} \hat{b}_{in},\\ \\ 
\frac{d \delta\hat{q}_{1} }{d t}=\Omega \delta \hat{p}_{1}, \\ \\
\frac{d \delta\hat{q}_{2} }{d t}=\Omega \delta \hat{p}_{2}, \\ \\
\frac{d \delta\hat{p}_{1} }{d t}=-\Omega \delta \hat{q}_{1}+g\left(|a_{s}|^{2}+a_{s}^{*}\delta\hat{a}+a_{s}\delta \hat{a}^{\dagger}\right), \\ \\
\frac{d \delta\hat{p}_{2} }{d t}=-\Omega \delta \hat{q}_{2}+g\left(|b_{s}|^{2}+b_{s}^{*}\delta\hat{b}+b_{s}\delta \hat{b}^{\dagger}\right).
\end{array}
\end{equation}
By further introducing the position and momentum quadratures for the two cavity modes and  their  input noises,
\begin{eqnarray}\label{lineareqquad}
\frac{d \delta\hat{X}_{a}}{d t}&=&\frac{d (\delta\hat{a}^{\dagger}+\delta\hat{a}) }{d t},\\ \nonumber 
\frac{d \delta\hat{P}_{a}}{d t}&=&i\frac{d (\delta\hat{a}^{\dagger}-\delta\hat{a}) }{d t},\\ \nonumber 
\frac{d \delta\hat{X}_{b}}{d t}&=&\frac{d (\delta\hat{b}^{\dagger}+\delta\hat{b}) }{d t},\\ \nonumber 
\frac{d \delta\hat{P}_{b} }{d t}&=&i\frac{d (\delta\hat{b}^{\dagger}-\delta\hat{b}) }{d t}, \\ \nonumber
\frac{d \delta\hat{X}_{in}^{a} }{d t}&=&\frac{d (\delta \hat{a}_{in}^{\dagger}+\delta \hat{a}_{in}) }{d t},\\ \nonumber 
\frac{d \delta\hat{P}_{in}^{a}}{d t}&=&i\frac{d (\delta \hat{a}_{in}^{\dagger}-\delta \hat{a}_{in}) }{d t},\\ \nonumber 
\frac{d \delta\hat{X}_{in}^{b} }{d t}&=&\frac{d (\delta \hat{b}_{in}^{\dagger}+\delta \hat{b}_{in}) }{d t},\\ \nonumber 
\frac{d \delta\hat{X}_{in}^{b}}{d t}&=&i\frac{d (\delta \hat{b}_{in}^{\dagger}-\delta \hat{b}_{in}) }{d t},
\end{eqnarray}
we can rewrite Eq.~\eqref{lineareq} in the  compact form
\begin{equation}\label{compct}
\frac{d R}{d t}=  {\mbox{\bf Z}} R + N.\\ 
\end{equation}
Here
\begin{equation}
\begin{array}{lll}
R^{T} & = & \big(\delta\hat{q}_{1},\delta\hat{p}_{1},\delta\hat{q}_{2},\delta\hat{p}_{2},\delta\hat{X}_{a}, \delta\hat{P}_{a}, \delta\hat{X}_{b}, \delta\hat{P}_{b}\big), \\ \\
N^{T} & = & \big(0,\epsilon_{1}(t),0,\epsilon_{2}(t),\sqrt{2 \kappa}\delta\hat{X}_{in}^{a}(t),\\ \\
& &   \sqrt{2 \kappa}\delta\hat{P}_{in}^{a}(t), \sqrt{2 \kappa}\delta\hat{X}_{in}^{b}(t), \sqrt{2 \kappa}\delta\hat{P}_{in}^{b}(t)\big),
\end{array}
\end{equation}
and
\begin{eqnarray*}
{\mbox{\bf Z}}&=&\left( \begin{array}{cc}
    {\mbox{\bf Z}_{1}} & {\mbox{\bf Z}_{2}} \\ 
  {\mbox{\bf Z}_{2}} &   {\mbox{\bf Z}_{3}}  \\ 
  \end{array} \right ),  \nonumber
\end{eqnarray*}
with the $4\times4$ matrices
\begin{equation}
\begin{array}{l}
  {\mbox{\bf Z}_{1}}=\left( \begin{array}{cccc}
    0 &  \Omega & 0 &0  \\ 
   -\Omega &-\gamma_{m}&0 & 0   \\ 
    0 & 0 & 0 &\Omega \\ 
  0 & 0 & -\Omega &-\gamma_{m}  \\ 
  \end{array} \right ),\\ \\ 
  {\mbox{\bf Z}_{2}}=\left( \begin{array}{cccc}
    0 &  0 & 0 &0  \\ 
  g_{a}^{s} &0&0 & 0   \\ 
    0 & 0 & 0 &0\\ 
  0 & 0 & g_{b}^{s} &0 \\ 
  \end{array} \right ), \\ \\ 
    {\mbox{\bf Z}_{3}}=\left( \begin{array}{cccc}
    -\kappa &  \Delta_{a} & 0 &\lambda  \\ 
   -\Delta_{a}&-\kappa&-\lambda & 0   \\ 
    0 & \lambda & -\kappa& \Delta_{b}\\ 
  -\lambda & 0 & -\Delta_{b} &-\kappa \\ 
  \end{array} \right ).
  \end{array}
  \end{equation}
The phase reference has been chosen such that $a_{s}$ and $b_{s}$ are real with $g_{a}^{s}=\sqrt{2}g a_{s}$ and $g_{b}^{s}=\sqrt{2}g b_{s}$. In Eq.~\eqref{compct}, ${\mbox{\bf Z}}$ is the drift matrix. Stability (in the steady state) demands that the real part of all the eigenvalues of ${\mbox{\bf Z}}$  must be negative. All the parameters in the present work have been chosen such that the system is 
stable in the steady state. 
\begin{figure}[!]
\centering
\includegraphics[width=0.451\textwidth]{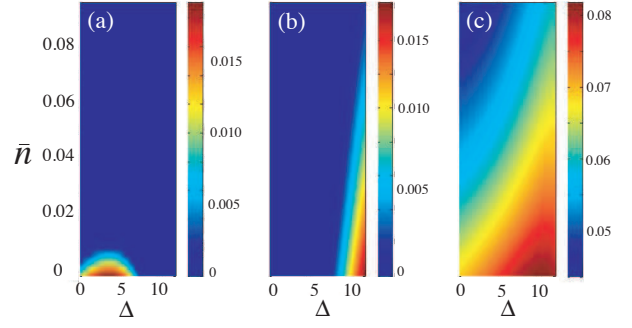}
\caption{\label{fig3}(Color online) Logarithmic  negativity as a measure of  entanglement between (a) two distant cavity mirrors, (b) a mirror and  adjacent cavity mode, and (c) a mirror and  distant cavity mode,  plotted  as a  function of  detuning $\Delta$ and average thermal occupancy of the two mirrors $\bar{n}_{1}=\bar{n}_{2}=\bar{n}$. We have chosen the different physical  parameters such that  $\Omega=1$, $g_{a}^{s}=g_{b}^{s}=2.5$, $\lambda=20$, $\kappa=0.08$, $\gamma_{m}=0.01$, and $\Delta_{a}=\Delta_{b}=\Delta$}
\end{figure} 

The dynamics of the coupled cavities with movable mirrors is governed by the first order matrix differential equation \eqref{compct}.  For an initial Gaussian state of the two cavities and their movable mirrors it is sufficient to fully characterize all the quantum correlations  by explicitly evaluating the  $8 \times 8 $ symmetric covariance matrix ${\mbox{\bf V}}$ where $V_{i,j}(t)=(\langle R_{i}(t)R_{j}(t)+R_{j}(t)R_{i}(t)\rangle)/2$. If the system is stable in the steady state, then the covariance matrix takes the form
\begin{equation}\label{newqeh}
V_{i,j}= \sum_{p,q} \int_0^{\infty}{ds} \int_0^{\infty}{ds'}W_{i,p}(s)W_{j,q}(s')\Phi_{p,q}(s-s'),  
\end{equation}
where ${\mbox{\bf W}}=\rm{exp}(\mbox{\bf Z} s)$ and $\Phi_{p,q}(s-s')=(\langle N_{p}(s)N_{q}(s')+N_{q}(s')N_{p}(s)\rangle)/2$ is the steady-state noise correlation matrix. It turns out that in the regime where  
the mechanical oscillators possess very high $Q$-values, the quantum Brownian noise becomes approximately $\delta$-correlated and in this limit the noise correlation matrix takes the form
\begin{equation}
\begin{array}{l}
\Phi_{p,q}(s-s')= \tilde{N}_{p,q} \delta(s-s'), \\ \\
\begin{array}{lll}
& = & \rm{Diag}[0,\gamma_{m}(2 \bar{n}_{1}+1)),0,\gamma_{m}(2\bar{n}_{2}+1),\\ \\
& & \kappa,\kappa,\kappa,\kappa] \delta(s-s'),
\end{array}
\end{array}
\end{equation}
with $\bar{n}_{1}=\left[e^{(\hbar \Omega/ k_{B} T_{1})}-1\right]^{-1}$ and $\bar{n}_{2}=\left[e^{(\hbar \Omega/ k_{B} T_{2})}-1\right]^{-1}$. Thus, Eq.~\eqref{newqeh} simplifies to
\begin{equation}
{\mbox{\bf V}}=\int_0^{\infty}{ds} {\mbox{\bf W}}(s)\tilde{{\mbox{\bf N}}}{\mbox{\bf W}}^{T}(s).
\end{equation}
When the system is stable in the steady state, the covariance matrix ${\mbox{\bf V}}$ satisfies a Lyapunov equation~\cite{smans3}
\begin{equation}\label{lyapug}
{\mbox{\bf Z}}{\mbox{\bf V}}+{\mbox{\bf V}} {\mbox{\bf Z}}^{T}=-\tilde{{\mbox{\bf N}}}.
\end{equation}

Once again, when we have the solution for the covariance matrix, we can compute various non-classical correlations between the optical and mechanical degrees of freedom. In particular, the degree of entanglement between different optical and mechanical modes can be evaluated by computing the logarithmic negativity as defined in \eqref{negtyeqn}. We numerically solve Eq.~\eqref{lyapug} for the covariance matrix $V$. An example of the numerically calculated logarithmic negativity between various optical and mechanical modes is presented in Fig.~\ref{fig3}. 

For evaluating the entanglement between various optical and mechanical modes we have chosen physical parameters accessible in present experiments. Not surprisingly, the steady-state  entanglement is  susceptible to thermal fluctuations of the environment.  A high temperature of the surrounding reservoirs will result in a completely separable state of the optical and mechanical modes.  One  should note that  the entanglement generated  between optical and mechanical modes in the steady state is not very large, but,  it does not require any 
quantum resources, such as additionally driving the two cavities with squeezed light~\cite{agarwal,laura}. 

Also, it is worth pointing out that with our particular choice of parameters we find that an appreciable entanglement appears between various  optical and mechanical modes only when we operate  far away from the regime of the red ($\Delta=\Omega$) or blue ($\Delta=-\Omega$) sideband. Although operating in the blue sideband regime is commonly considered ideal for generating entanglement between various optical and mechanical modes, the condition that the steady state should be stable puts serious restrictions on the coupling strength between the mechanical and optical modes~\cite{dvittwo}.

A challenging aspect of any scheme involving entanglement generation between "macroscopic" mechanical systems is the actual experimental detection of entanglement. There are however some recent promising proposals to create and detect quantum correlations in optomechanical settings~\cite{laura, mfent}. Since it is comparatively easier to detect quantum correlations between 
optical modes, as compared to directly detecting quantum entanglement between mechanical modes, the  essence of these proposals is to swap the nonlocal correlations from the mechanical modes back to the optical modes. As shown in Fig.~\ref{figzero} this can, for instance, be implemented using two auxiliary light modes, each initially prepared in classical uncorrelated states.  These auxiliary modes can be two modes of distant cavities, and the geometry so arranged that each entangled mirror couples independently to the two modes. The non-local correlations may then be transferred from the movable mirrors to the initially uncorrelated auxiliary modes, which may eventually become entangled.  Thus, using standard homodyne measurement techniques, the entire correlation matrix of the two optical  auxiliary modes can be reconstructed. A presence of non-zero 
quantum correlations between these optical modes will 
be an indirect signature of non-zero quantum correlations  between the mechanical modes. 

\section{Conclusion}
\label{sec:conclude}
We have discussed in detail the possibility of generating nonlocal quantum correlations between optical and  mechanical modes of two spatially separated  cavities. Each cavity  is assumed to have one fixed and one movable mirror and the two cavities are coupled by an optical fiber. Under the Born-Oppenheimer approximation, relying on separating the dynamics into a fast optical timescale and a slow mechanical timescale,  we have analytically worked out the dynamics of the two coupled movable mirrors. Furthermore, within this adiabatic regime, we have also presented a full analytical solution of the master equation governing the open-system  dynamics of the two movable mirrors. We especially found that  the interaction mediated via two optical modes entangles the two distant movable mirrors. Cavity losses were taken into account in an effective non-Hermitian model, and mirror entanglement was found to be fairly robust against such dissipation.
Using a complementary model, we have studied the 
two coupled cavities using the quantum  Langevin formalism, by explicitly solving the resulting equations of motion. In the presence of strong driving laser fields we have found that the two coupled cavities exhibit nonlocal quantum correlations between distant optical and mechanical modes. In particular, these optical and mechanical modes exhibit entanglement in the steady state and at finite temperatures.  This  opens up  an interesting possibility to study spatially separated massive  Schr\"odinger cat states.

During the completion of this work we became aware of a recent proposal  discussing the possibility of phonon photon entanglement in coupled optomechanical arrays \cite{akram2}.

\section{Acknowledgements} We gratefully acknowledge fruitful discussions with U.~Akram and G.~J.~Milburn.  C.J. acknowledges Michael Hall for introducing him to an alternative way to solve the master equation and also support from the ORS scheme. J.L. acknowledges financial support from the Swedish Research Council (VR), Deutscher Akademischer Austausch Dienst (DAAD), and Kungl. Vetenskapsakademien (KVA), M.J. support from the Swedish Research Council (VR) and the Korean WCU program funded by MEST/KOSEF (R31-2008-000-10057-0), and E.A. acknowledges support from EPSRC  EP/G009821/1.


\begin{thebibliography}{99}
\bibitem{coherent} This follows from noting that an initial coherent state with amplitude $\alpha$ has the quantum characteristic function $\chi(\epsilon)=e^{\epsilon \alpha^{*}-\rm{c.c.}}$.  Under purely dissipative time evolution $\chi$ satisfies  $\frac{\partial \chi}{\partial t}=-\kappa (\epsilon^{*}\frac{\partial }{\partial \epsilon^{*}}+\rm{c.c.})\chi$, the solution of which is $\chi(\epsilon)=e^{\epsilon \alpha^{*}e^{-\kappa t}-\rm{c.c.}}$.
\bibitem{combath} Assuming independent baths implies neglecting any reservoir-induced correlations, which is supposed to be justified in the present setup of spatially separated oscillators.
\end{thebibliography}
 \end{document}